\documentclass[10pt,conference]{IEEEtran}

\def\BibTeX{{\rm B\kern-.05em{\sc i\kern-.025em b}\kern-.08em
    T\kern-.1667em\lower.7ex\hbox{E}\kern-.125emX}}
    
\usepackage[sort,compress]{cite}
\usepackage{amsmath,amssymb,amsfonts}
\usepackage{mathtools}
\usepackage{stmaryrd}
\usepackage{algorithmic}
\usepackage{graphicx}
\usepackage{textcomp}
\usepackage{xcolor}
\usepackage{bm}
\usepackage{adjustbox}
\usepackage[hidelinks]{hyperref}
\usepackage{tabularx}
\usepackage{float}
\usepackage{colortbl}
\usepackage{booktabs}
\usepackage{listings}
\usepackage{multirow}
\usepackage{graphicx}
\usepackage{xspace}
\usepackage[most]{tcolorbox}
\usepackage{balance}
\usepackage[normalem]{ulem}
\usepackage{tabularray}
\usepackage{pgfplots}
\usepackage{threeparttable}
\usepackage{tikz}
\usepackage[shortlabels]{enumitem}
\usepackage{pifont}
\usepackage{pgfplots}
\usetikzlibrary{pgfplots.groupplots}
\usepackage{inconsolata}
\usepackage{ulem}
\usepackage{fontawesome}

\definecolor{rbrtext}{RGB}{197,53,152}
\definecolor{icetext}{RGB}{6,52,200}
\definecolor{bctext}{RGB}{36,128,20}
\definecolor{ccpbg}{RGB}{253,230,158}
\definecolor{pspbg}{RGB}{199,224,182}
\definecolor{eppbg}{RGB}{219,219,219}
\definecolor{opbg}{RGB}{182,199,230}

\newcommand{\coloruline}[2][black]{{\textcolor{#1}{\uline{\textcolor{black}{#2}}}}}

\pgfplotsset{compat=1.18}
\useunder{\uline}{\ul}{}

\newcommand{\greyboxb}[2]{
\vspace{0.05cm}
    \begin{tcolorbox}[
        left=2pt, right=2pt, top=2pt, bottom=2pt,
        boxrule=0.2mm,
        leftrule=2mm,
        arc=0mm,
        colframe=black!40!white, 
        colback=black!5!white, 
        colbacktitle=black!50!white 
    ]
    \textbf{#1}{#2}
    \end{tcolorbox}
\vspace{0.05cm}
}

\newcommand{\phead}[1]{\vspace{1mm} \noindent {\bf #1}}

\newcommand{\newname}{$\bm{\mathcal{R}}$\textbf{E}val\xspace}
\newcommand{\normnewname}{REval\space}

\newcommand{\newidea}{Incremental Consistency\xspace}
\newcommand{\abbrnewidea}{IC\xspace}

\begin{document}

\title{Reasoning Runtime Behavior of a Program with LLM: How Far Are We?}

\author{}
\author{
\IEEEauthorblockN{Junkai Chen\IEEEauthorrefmark{1}}
\IEEEauthorblockA{
\textit{School of Software Technology,} \\
\textit{Zhejiang University} \\
Ningbo, China \\
junkaichen@zju.edu.cn}
\\\\
\IEEEauthorblockN{Zhenhao Li}
\IEEEauthorblockA{
\textit{York University} \\
Toronto, Canada \\
zhenhao.li@ieee.org}
\and
\IEEEauthorblockN{Zhiyuan Pan\IEEEauthorrefmark{1}}
\IEEEauthorblockA{
\textit{The State Key Laboratory of} \\
\textit{Blockchain and Data Security,} \\
\textit{Zhejiang University} \\
Hangzhou, China \\
zy\_pan@zju.edu.cn}
\\
\IEEEauthorblockN{Ge Li}
\IEEEauthorblockA{
\textit{Peking University}\\
Beijing, China \\
lige@pku.edu.cn}
\and
\IEEEauthorblockN{Xing Hu\IEEEauthorrefmark{2}}
\IEEEauthorblockA{
\textit{The State Key Laboratory of} \\
\textit{Blockchain and Data Security,} \\
\textit{Zhejiang University}\\
Hangzhou, China \\
xinghu@zju.edu.cn}
\\
\IEEEauthorblockN{Xin Xia}
\IEEEauthorblockA{
\textit{Zhejiang University}\\
Hangzhou, China \\
xin.xia@acm.org}
}

\maketitle

\begingroup\renewcommand\thefootnote{\IEEEauthorrefmark{1}}
\makeatletter\def\Hy@Warning#1{}\makeatother
\footnotetext{~Equal contribution.}
\endgroup

\makeatletter\def\Hy@Warning#1{}\makeatother
\begingroup\renewcommand\thefootnote{\IEEEauthorrefmark{2}}
\footnotetext{~Corresponding author.}
\endgroup

\begin{abstract}
Large language models for code (i.e., code LLMs) have shown strong code understanding and generation capabilities. To evaluate the capabilities of code LLMs in various aspects, many benchmarks have been proposed (e.g., HumanEval and ClassEval). 
Code reasoning is one of the most essential abilities of code LLMs (i.e., predicting code execution behaviors such as program output and execution path), but existing benchmarks for code reasoning are not sufficient. 
Typically, they focus on predicting the input and output of a program, ignoring the evaluation of the intermediate behavior during program execution, as well as the logical consistency (e.g., the model should not give the correct output if the prediction of execution path is wrong) when performing the reasoning.  
To address these problems, in this paper, we propose a framework, namely \newname, for evaluating code reasoning abilities and consistency of code LLMs with program execution. 
We utilize existing code benchmarks and adapt them to new benchmarks within our framework. A large-scale empirical study is conducted and most LLMs show unsatisfactory performance on both Runtime Behavior Reasoning (i.e., an average accuracy of 44.4\%) and Incremental Consistency Evaluation (i.e., an average IC score of 10.3). 
Evaluation results of current code LLMs reflect the urgent need for the community to strengthen the code reasoning capability of code LLMs. 
Our code, data and \newname leaderboard are available at \url{https://r-eval.github.io}.
\end{abstract}


\section{Introduction}
\label{sec:intro}

Large language models (LLMs) attract great attention for their exceptional performance on diverse tasks~\cite{min2023recent} including sentiment analysis~\cite{barbieri2022xlm}, logical reasoning~\cite{creswell2022selection}, and question answering~\cite{rogers2023qa}. 
Recently, large language models for code (i.e., code LLMs) have become a popular research area because of the promising prospect of empowering humans in software development and maintenance~\cite{fan2023large}. Hence, both academia and industry have proposed a lot of code LLMs (e.g., CodeLlama family~\cite{codellama} and Magicoder series~\cite{magicoder}), which are widely applied to different tasks like code generation~\cite{codex, mbpp}.

To provide a fair and comprehensive measure of the capabilities of code LLMs, many code-related benchmarks (e.g., HumanEval~\cite{codex} and CodeXGLUE~\cite{codexglue}) are proposed to evaluate the effectiveness of code LLMs in different tasks such as code generation and vulnerability detection~\cite{chang2023survey}. 
Given that ``executable'' is a distinct feature of code compared to natural language, and code execution provides additional information (e.g., program output) to assist with code tasks~\cite{ni2023lever,chen2023teaching}, benchmarking code reasoning abilities of code models with execution raises researchers' interests~\cite{gu2024cruxeval,liu2024codemind}. 
Here, code reasoning is referred to as predicting code execution behaviors (e.g., program outputs, execution paths and possible variable values) without executing the code directly.
For example, Gu et al.~\cite{gu2024cruxeval} proposed CRUXEval to evaluate code LLMs by predicting output from input and vice versa. 
Typically, these works measure the model's ability to predict and analyze the relationship between the input and output of an executable program. However, the intermediate information (e.g., execution path) \textit{during} code execution is ignored, posing challenges to developers in comprehending the program's runtime behavior.

\begin{figure}[t]
\centerline{\includegraphics[width=1.0\linewidth]{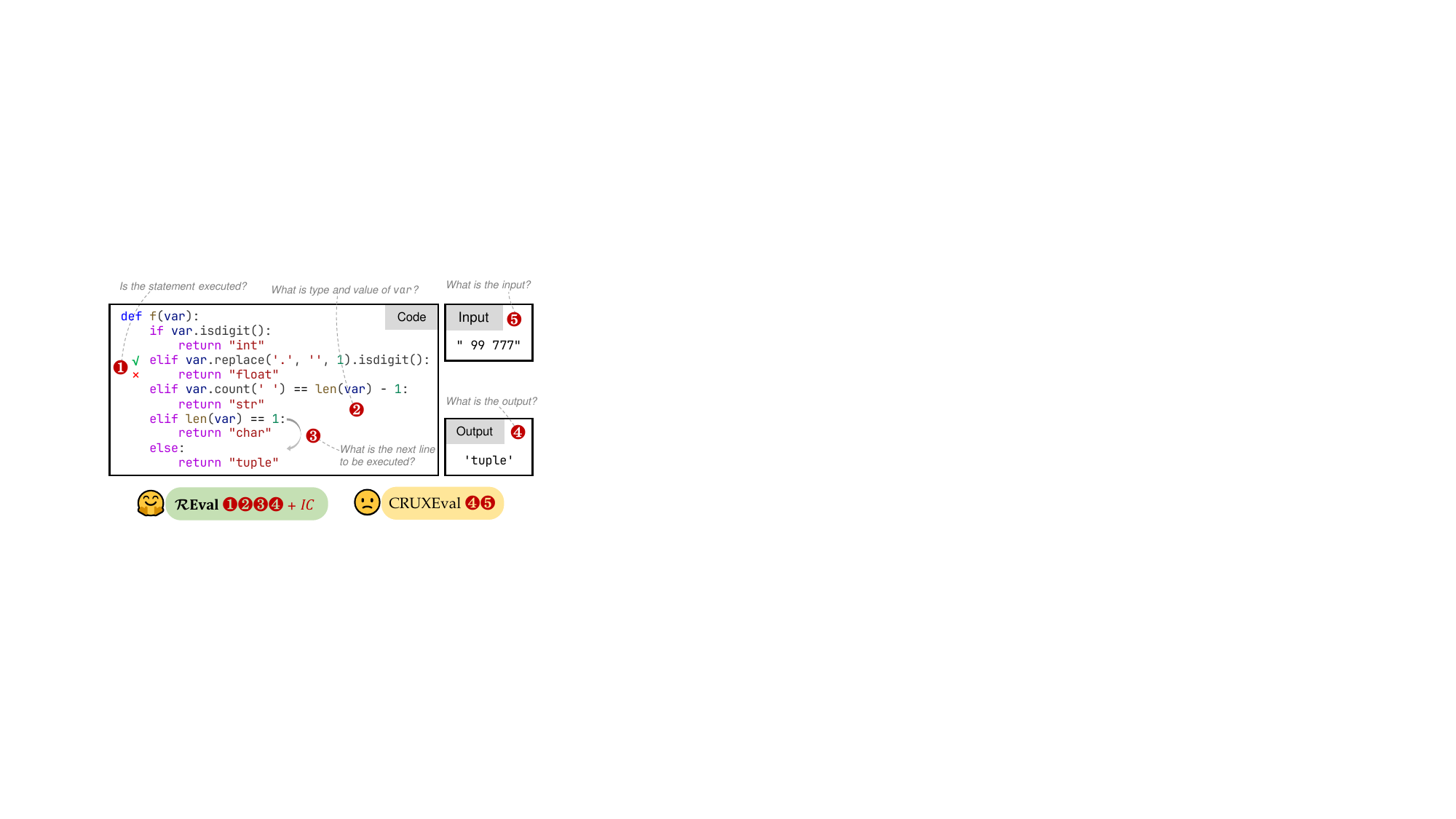}}
\caption{
The demonstration of code reasoning tasks in CRUXEval~\cite{gu2024cruxeval} and \newname. ``\abbrnewidea'': \newidea.
}
\label{fig:comparsion}
\end{figure}

Fig.~\ref{fig:comparsion} shows common concerns about the \textit{runtime behavior} during program execution. 
Intuitively, how a program behaves under certain input can help developers better understand the code and perform debugging activities. 
For example, if we have concerns about the correctness of a certain statement while debugging a snippet of code, we typically first determine whether this statement is executed given the input (i.e., ~\ding{182} in Fig.~\ref{fig:comparsion}); 
If it is executed, observing the changes in variables before and after execution is a natural choice~\ding{183}; 
Sometimes this line of code may seem fine, so the statement immediately following it will be examined~\ding{184}; 
Additionally, the program output can be used to verify whether the results match the expectations~\ding{185}. 
Therefore, we argue that these kinds of runtime behaviors (e.g., program state and execution path) are essential for program understanding and reasoning for humans. Meanwhile, they are also proven to be effective for an in-depth understanding of code semantics for language models~\cite{ding2024traced}. 
As previous benchmarks like CRUXEval (i.e., with~\ding{185} and \ding{186}) fail to evaluate whether LLMs can reason about these dynamic characteristics of a program, it is necessary to measure the code reasoning ability of LLMs with runtime behavior of execution. 
In this paper, we propose our framework, \newname, to comp\underline{\textbf{re}}hensively (\underline{\textbf{re}})-evaluate the code \underline{\textbf{re}}asoning ability of LLMs, which consists of two evaluation components: (i) Runtime Behavior Reasoning and (ii) \newidea Evaluation.

\noindent \textbf{Evaluation Component 1: \textit{Runtime Behavior Reasoning}.} To mitigate this limitation in previous research, we make the first attempt to systematically evaluate the code LLM's ability to reason about the runtime behavior of program execution. Specifically, we propose four evaluation tasks to achieve this goal:
\ding{182} Code Coverage Prediction (CCP), i.e., whether a statement is executed or not; 
\ding{183} Program State Prediction (PSP), i.e., what is the value and the type of a variable; 
\ding{184} Execution Path Prediction (EPP), i.e., which is the next statement to be executed; 
and \ding{185} Output Prediction (OP), i.e., what is the output. 
These four tasks cover various aspects of program execution, including control flow, data flow, and type dependency, which are widely applied to prior research in software engineering such as type inference~\cite{peng2022static} and code translation~\cite{guo2020graphcodebert}. 
Therefore, this evaluation provides a more comprehensive measure of code model's ability to reason about executable programs in comparison with previous work. 

Nevertheless, it is noticed that sometimes the reasoning results of a model could conflict with human logic on \textit{sequential} tasks in Runtime Behavior Reasoning. 
For instance, the code model may correctly predict the next statement to be executed (i.e., EPP) when it fails to tell the value of a variable after the statement's execution (i.e., PSP), which is not expected because the control flow of the execution relies on the program state. 
As this kind of \textit{inconsistency in sequentially related tasks} is unlikely to occur in humans, the trustworthiness of AI systems built on these models (e.g., GitHub Copilot~\cite{copilot}) can easily suffer from these unreliable behaviors. 
Although some previous works have discussed consistency for code LLMs~\cite{allamanis2024unsupervised,min2023beyond}, they are limited to semantic consistency like back translation between NL and code and ignore the logical consistency mentioned here. 
Hence, it is necessary to measure the consistency of code LLMs on sequentially related tasks.

\noindent \textbf{Evaluation Component 2: \textit{\newidea Evaluation}.} To fill the gap in evaluation, we propose a novel metric named \newidea (IC) to measure the extent to which the model can maintain its logical consistency on sequentially related tasks of \textit{incremental} difficulty. 
We observe that the four tasks in Runtime Behavior Reasoning are progressive and consistent with the context of IC, i.e., the knowledge required to finish the current task is the preliminary of the next task. 
Hence, we can judge how much a model is incrementally consistent by utilizing the results of reasoning runtime behavior (See Section~\ref{sec:framework} for details). Incremental Consistency provides new sights for evaluating LLMs and the consistency measure of AI systems beyond traditional metrics. 

To construct our framework, we leverage existing executable datasets (e.g., HumanEval~\cite{codex} and ClassEval~\cite{classeval}) as our base benchmarks and adapt them into an adapted benchmark within our framework by extracting runtime behavior, constructing, and filtering the problems. 
We conduct a large-scale empirical study on various models, including general and code LLMs in our frameworks. Evaluation results show that our framework presents a degree of difficulty and most LLMs show poor performance on both Runtime Behavior Reasoning and IC Evaluation (e.g., an IC score below 20 for all open-source LLMs we evaluate). Our research highlights the importance of utilizing runtime behavior and incremental consistency evaluation to measure the reasoning ability of code LLMs, and we call for targeted efforts in subsequent research to enhance these weaknesses.

In summary, the contributions of our paper are as follows:
\begin{itemize}[leftmargin=*]
    \item We propose a new framework, \newname, to comprehensively evaluate code LLMs' abilities of code reasoning. 
    To the best of our knowledge, we are the first work to evaluate code models to systematically reason about runtime behavior during program execution.
    \item We propose a novel metric named \newidea (\abbrnewidea) to measure to what extent a code LLM can maintain its consistency across sequentially related tasks of incremental difficulty.
    \item We conduct a large-scale empirical study on diverse LLMs within our evaluation framework. 
    Our results reveal the limitations of reasoning runtime behavior and IC of code models.
    \item We construct an adapted benchmark based on HumanEval~\cite{codex} and ClassEval~\cite{classeval} and develop an evaluation harness for our framework. 
    To facilitate further research of code reasoning, our code, data, and \newname leaderboard are publicly available at \url{https://r-eval.github.io}.
\end{itemize}

\section{Background and Related Work}
\label{sec:background}

In this section, we discuss the background information of our research and the corresponding related work.

\subsection{Code Execution and Reasoning}

\subsubsection{Code Execution Behavior}

We refer to code execution behavior as the additional information offered by program execution compared to static analysis. According to the execution order, we classify them into pre/post-execution information and runtime information: 
\begin{itemize}
\item \textit{Pre/Post-Execution Information} is the content we can obtain before or after the actual execution process of program. For example, the input, output, and NL requirements belong to this category. 

\item \textit{Runtime Information} is the intermediate state during code execution. For instance, we are able to collect contents like program state and execution path only when the code is still running.
\end{itemize}

Previous research has leveraged code execution behavior to improve the performance of various downstream tasks, e.g., program understanding~\cite{code_vectors}, code generation~\cite{chen2023teaching,ni2023lever,liu2023code}, software testing~\cite{test_execution_traces,exectution_trace_embedding} and vulnerability detection~\cite{ding2024traced}. 
Ni et al.~\cite{ni2023lever} improved code generation performance with an extra verifier, which learns the results of code execution and helps rerank generated code candidates. 
Chen et al.~\cite{chen2023teaching} utilized different kinds of feedback including output to help LLMs ``self-debug'' the generated code. They designed a series of prompting strategies to guide the model to refine the program automatically. 
In these works, pre-/post-execution information, such as program output, is applied to code generation. Furthermore, some works found the worth of dynamic features during execution and exploit them to train various language models. 
Liu et al.~\cite{liu2023code} pre-trained a language model to learn the execution process of the program. Specifically, they represented the program state as a sequence that neural models can learn from and expect the model to predict the trace. 
Compared to Liu et al., Ding et al.~\cite{ding2024traced} proposed a pre-training technique combining both static and dynamic characteristics of the program. 
In summary, the aforementioned works reflect the close relationship between the behavior of code execution and the program semantics, and emphasize the importance of evaluating models for code reasoning with execution.

\subsubsection{Code Reasoning with Large Language Models} 

As introduced in Section~\ref{sec:intro}, in the task of code reasoning, an LLM needs to predict the program behavior without execution.

Recently, some works have proposed different evaluation approaches for the code reasoning abilities of code LLMs. 
For example, Gu et al.~\cite{gu2024cruxeval} proposed CRUXEval, which requires LLMs to reason about pre/post-execution information such as input and output. Following this study, similar to the idea of CRUXEval, Liu et al.~\cite{liu2024codemind} extended the evaluation tasks (i.e., predicting input and output) to the natural language specification. 
However, their evaluation approaches are still limited to pre/post-execution information and ignore intermediate runtime behavior. In contrast, our work goes a step further to measure how the model learns the runtime behavior \textit{during} execution, which shows promising potential in helping program comprehension, as mentioned above.
We notice that a recent work~\cite{la2024code} aimed to simulate the code execution process with code LLMs. They used the analogy of a large language model to a CPU to explore the process of a program executing code, paying more attention to algorithm complexity and structure. 
Different from the aforementioned studies, our framework is not only limited to algorithm problems (e.g., competition-level ones), but also suitable for general programming scenarios (e.g., more real-world projects). In addition, we also explore detailed runtime behavior like code coverage and execution path, containing more runtime information.

\subsection{Consistency for Large Language Models}


\phead{Semantic Consistency.} 
Semantic consistency refers to the same decisions on semantically equivalent texts of LLMs~\cite{jang2022becel}. For example, the model should provide similar and even the same answers in the face of two meaning-perserving questions. 
In the realm of software engineering, this feature is generally utilized for the unsupervised evaluation of code LLMs~\cite{min2023beyond,allamanis2024unsupervised}: 
Min et al.~\cite{min2023beyond} evaluated the self-consistency of code LLMs by comparing the functional correctness of two code snippets: one code is generated using a human-written description, and the other is generated iteratively with the summary of the previously generated. 
Chen et al.~\cite{chen2024nlperturbator} studied the robustness of code LLMs to the variations in natural language descriptions for code generation.
Allamanis et al.~\cite{allamanis2024unsupervised} introduced round-trip correctness which aligns code and NL to perform unsupervised evaluation for code LLMs. 
The aformentioned works leveraged the back translation between NL and PL iteratively generated by the model and conduct the semantic or functional comparison. However, they are restricted to the context of semantic consistency in the context of NL and PL.

\phead{Logical Consistency.} 
If an LLM is able to make predictions without logical contradiction, it shows the feature of logical consistency~\cite{jang2022becel}. For example, if one model assumes a proposition to be true, it should consider the negation of that proposition to be false as well. 
There are lots of previous research about how to evaluate and utilize logical consistency for LLMs in natural language processing~\cite{jang2023consistency,jang2022becel,elazar2021measuring,sahu2022unpacking}, but few works pay attention to logical consistency on code LLMs. As the reasoning ability is highly related to its logical consistency~\cite{wang2022self}, a comprehensive code reasoning evaluation should contains the measure of logical consistency in scope of programming languages (PLs). Therefore, it motivates us to propose a novel consistency metric idea named IC to fill this gap.

\begin{figure*}[t]
\centerline{\includegraphics[width=1.0\linewidth]{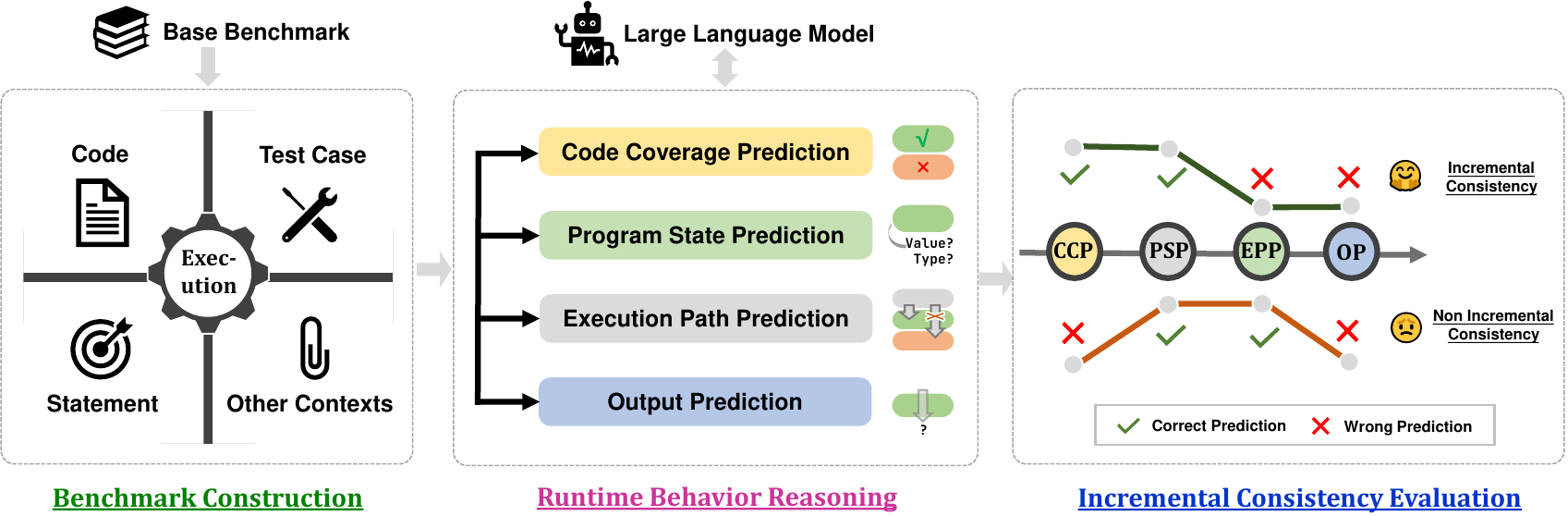}}
\caption{
Overview of our framework. 
\textbf{Benchmark Construction}: we adapt the base benchmarks to fit our framework by execution. 
\textbf{Runtime Behavior Reasoning}: we propose four tasks including CCP, PSP, EPP, and OP, which challenge LLMs to perform code reasoning. 
\textbf{Incremental Consistency Evaluation}: we evaluate if the model can maintain consistency on sequentially related tasks (i.e., \newidea).
}
\label{fig:framework}
\end{figure*}

\subsection{Code LLMs and Benchmarks}
 \subsubsection{Code LLMs}

Large language models for code are LLMs specialized for the generation and understanding of PLs. For example, 
CodeLlama family models~\cite{codellama} inherit the architecture of Llama2~\cite{llama2} and are further pre-trained on extra code corpora. Its three variant models (i.e., {\em base}, {\em instruct}, and {\em Python-specialized}) are designed for different programming scenarios.
StarCoder2~\cite{starcoder2} is a series of code LLMs developed by the BigCode Project, which achieves competitive performance with other similar-sized models. These models are trained on The Stack v2 dataset~\cite{starcoder2} whose data size is four times larger than its first generation. 
CodeGen2.5 models~\cite{codegen25} are improved versions of their previous models (e.g., CodeGen2) for program synthesis. It is claimed that their performance gains mainly come from optimizations such as training and sampling strategies.

Code reasoning is one of the most important capabilities of LLMs related to code~\cite{liu2024codemind,gu2024cruxeval}, but few work are dedicated to its evaluation. In this paper, we aim at comprehensively evaluating their reasoning capabilities for programming languages based on execution.

\subsubsection{Benchmarks}

Recently, many code generation benchmarks have been proposed to evaluate the correctness of code snippets generated by code LLMs.
HumanEval~\cite{codex} is one of the most popular benchmarks for code generation. It consists of 164 competitive programming problems and evaluates the functional correctness of generated samples rather than text similarity. 
Apart from competitive benchmarks of which question generally runs in a simple context, lots of context-aware benchmarks such as CoderEval~\cite{codereval}, ClassEval~\cite{classeval}, and CrossCodeEval~\cite{ding2024crosscodeeval} have been proposed. These benchmarks provide more complex surrounding contexts and dependencies (e.g., private libraries) to evaluate code generation in real-world projects.
In addition, some domain-specific benchmarks have been proposed to evaluate the performance of code generation in various programming languages and paradigms (e.g., DS-1000~\cite{ds1000} for data science). 
Our work utilizes existing executable benchmarks for code generation and evaluates code LLMs with respect to code reasoning, which makes our framework universal and applicable to different scenarios.

 Prior studies proposed a variety of code understanding tasks including code search, type inference, and code translation. 
 Evaluating the ability of code understanding is essential for code LLMs. 
Lu et al.~\cite{codexglue} propose CodeXGLUE, a comprehensive benchmark for code models that supports 10 tasks related to code and text. 
Cassano et al.~\cite{caniedit} create a hand-crafted benchmark to evaluate the instruction following ability on code editing. 
Khan et al.~\cite{xcodeeval} introduce a large-scale multilingual multitask benchmark that consists of numerous executable coding examples. 
Compared to the above benchmarks for code understanding, we propose a comprehensive framework for code reasoning from the perspective of runtime behavior, which provides a different point of view for the evaluation of code LLMs.

\section{\normnewname Framework}
\label{sec:framework}

In this section, we first introduce the overview of our evaluation framework \newname, and then describe the two evaluation components in detail, namely, \textit{\textbf{Runtime Behavior Reasoning}} and \textit{\textbf{Incremental Consistency Evaluation}}. In the end, we describe how to construct the corresponding benchmark under our framework.

\subsection{Overview of Framework}
Fig.~\ref{fig:framework} shows an overview of our framework, which aims to challenge code LLMs to reason how the program behaves during execution.
To achieve this, we adopt two different perspectives for the abilities of code reasoning: (1) \textit{Runtime Behavior Reasoning}; and (2) \textit{Incremental Consistency Evaluation}. 

For Runtime Behavior Reasoning, we focus on whether the code model can correctly predict the intermediate states of program execution given an executable program and input (as well as other contexts in the base benchmark). We select four different dimensions of runtime behavior, including \textit{code coverage}, \textit{execution path}, \textit{program state}, and \textit{output}, each of which corresponds to a specific sub-task under Runtime Behavior Reasoning. We present the task description and the evaluation metric in Section~\ref{subsec:runtime} for each sub-task.

Besides the standalone metrics that measure a single capability of the models, we propose a novel idea named~\newidea Evaluation to assess the consistency across a series of incremental tasks during code reasoning. 
As the knowledge required to finish the latter task contains that of the former task in Runtime Behavior Reasoning, the difficulty increases progressively in order, and we can utilize this characteristic to evaluate \newidea of LLMs with existing predictions (see Section~\ref{subsec:incremental} for details).

In addition, as our evaluation relies on existing base benchmarks, we present how to construct an adapted benchmark of code reasoning within our framework in Section~\ref{subsec:benchmark}.

\subsection{\coloruline[rbrtext]{Runtime Behavior Reasoning}}
\label{subsec:runtime}

Runtime behavior refers to the intermediate state and information during program execution, such as code coverage and variable values, which are widely mentioned in previous research~\cite{ding2024traced,tufano2023predicting}. 
As shown in Fig.~\ref{fig:framework}, to evaluate the reasoning ability in program runtime behavior for code models, we analyze and select four representative dimensions of intermediate information during execution, including \textit{code coverage}, \textit{execution path}, \textit{program state}, and \textit{output prediction}. 
Corresponding to these features, we propose four sub-tasks for Runtime Behavior Reasoning and introduce them in detail.

\subsubsection{\coloruline[ccpbg]{Code Coverage Prediction (CCP)}}
Code coverage measures the proportion of code covered by a test suite~\cite{hemmati2015effective}. Recent research~\cite{tufano2023predicting} utilized this idea to challenge the model in predicting whether the statements in the program can be executed or not. Hence, we exploit the LLM code to judge whether a specific statement is executed given the input of a test case. 

\phead{Task Description.} 
Given a program $\mathbf{P}$ with statements $(S_1, S_2, \cdots, S_n)$, an input $\mathbf{X}$ for execution and a statement index $I$, the model $\mathcal{M}$ is required to predict whether the $I$-th statement $S_I$ is executed. The ground truth can be denoted as $\text{Coverage}(I)$.

\phead{Evaluation Metrics.} 
In this task, \textbf{Accuracy} presents the percentage of correct coverage predictions. For our benchmark that consists of $N$ number of $(\mathbf{P}, \mathbf{X}, I)$ pairs, \textit{Accuracy} can be computed as:
$$ \text{Accuracy} = \displaystyle\dfrac{1}{N}\sum \overbracket[0.5pt][8pt]{\left\llbracket \mathcal{M}(\mathbf{P}, \mathbf{X}, I) = \text{Coverage}(I) \right\rrbracket}^{\text{Conditional Expression. Value is 0 or 1.}} $$

Besides, as this task can be considered as a binary classification task, we also use \textbf{F1 score} as the evaluation metric following previous research~\cite{liu2023code}.


\subsubsection{\coloruline[pspbg]{Program State Prediction (PSP)}} 
The initial idea of \textit{program state} refers to values of the program counter and the variables~\cite{ammann2016introduction} in the context of assembly language and instructions. 
Since we are mainly concerned with code models at the source code level, we follow related work~\cite{ding2024traced,souza2023lexecutor} and define \textit{program state} as a set of variables in the current runtime scope. Each variable has its corresponding value and type. 
Program State Prediction examines the model's ability to reason about value and type conversion of the variable after a statement is executed.

\phead{Task Description.} 
Given a program $\mathbf{P}$ with statements $(S_1, S_2, \cdots, S_n)$, an input $\mathbf{X}$, a statement index $I$ and a variable name $V$ related to the current statement $S_i$, the model $\mathcal{M}$ is required to predict the type and value of variable $V$ after $S_I$ is executed. The ground truth of type and value can be denoted as $\text{Ty}(I, V)$ and $\text{Val}(I, V)$, respectively.

\phead{Evaluation Metrics.} 
In this task, \textbf{Accuracy} (Acc.) measures the percentage of correct value and type predictions:

$$
    \text{Acc.} = \displaystyle\dfrac{1}{N}\displaystyle\sum \llbracket \mathcal{M}(\mathbf{P}, \mathbf{X}, I, V) = (\text{Val}(I, V), \text{Ty}(I, V))\rrbracket 
$$

With this equation, a model's prediction is correct only if the value and type both match the ground truth.


\subsubsection{\coloruline[eppbg]{Execution Path Prediction (EPP)}} 
In this task, we refer to the execution path of the program as ordered sequences of statements. As the granularity of our context is statement-level, we challenge the code model to predict the next statement to be executed given a specific statement. 
A code LLM skilled in code reasoning should be capable of telling where the control flow of the program is going and, consequently, can predict the next executed statement naturally.

\phead{Task Description.}
Given a program $\mathbf{P}$ with statements $(S_1, S_2, \cdots, S_n)$, an input $\mathbf{X}$ and a statement index $I$, the model $\mathcal{M}$ is required to predict the next statement to be executed after $S_I$ is executed. The ground truth can be denoted as $\text{Next}(I)$.

\phead{Evaluation Metrics.} 
In this task, \textbf{Accuracy} measures the percentage of correct next statement predictions:

$$ \text{Accuracy} = \displaystyle\dfrac{1}{N}\sum \left\llbracket \mathcal{M}(\mathbf{P}, \mathbf{X}, I) = \text{Next}(I) \right\rrbracket $$

Note that if the number of possible answers is more than one (i.e., several statements could be the next one to execute), we consider the prediction correct if it hits any possible one.

\subsubsection{\coloruline[opbg]{Output Prediction (OP)}} 
This task is to directly generate the output of a program with the given input, which is applied in previous code reasoning work~\cite{gu2024cruxeval,liu2024codemind}. 
To accurately predict the output of a program, code LLMs should be capable of controlling and simulating the whole execution process, which places high demands on the code reasoning ability. To evaluate the correctness of the output, we utilize test cases (i.e., a collection of assertion statements) in the base benchmarks. This approach is also applied in various code benchmarks~\cite{codex,classeval}.

\phead{Task Description.}
Given a program $\mathbf{P}$ and an input $\mathbf{X}$ for execution, the model $\mathcal{M}$ is required to generate the output. The correct output can be denoted as $\mathbf{Y}$.

\phead{Evaluation Metrics.}
In this task, \textbf{Accuracy} measures the percentage of correct output predictions:

$$ \text{Accuracy} = \displaystyle\dfrac{1}{N'}\sum \left\llbracket \mathcal{M}(\mathbf{P}, \mathbf{X}) = \mathbf{Y} \right\rrbracket,$$
where $N'$ equals the number of different $(\mathbf{P}, \mathbf{X})$ pairs in our benchmark.


\subsection{\coloruline[icetext]{Incremental Consistency Evaluation}}
\label{subsec:incremental}

\newidea refers to the idea of how much the model can maintain its consistency across a series of sequentially related tasks. Intuitively, if an LLM cannot reason about the current task, it is not expected to finish the next task whose preliminary depends on the current task. To clarify this idea, we first present the description of \newidea, and then explain how we evaluate it on code models in practice.

\subsubsection{Description of \newidea} 

The core idea of \newidea is to assess code models by leveraging the relationship where the knowledge from one task in a series of tasks depends on the next task. In the context of our research, four distinct sub-tasks (i.e., CCP, PSP, EPP, and OP) in Runtime Behavior Reasoning are selected, and we can observe some patterns from them: 
\begin{enumerate}[i)]
\item $\text{CPP}\Leftarrow\text{PSP}$: Since the execution of a statement could lead to changes in the program state, the prerequisite for PSP is correctly predicting if the statement is executed (i.e., CCP). 
\item $\text{PSP}\Leftarrow\text{EPP}$: The control flow of a running program is affected by the value of some variables (e.g., ``if" branch and its conditional variable), thus the next statement to be executed (i.e., EPP) is influenced by the program state (i.e., PSP).
\item $\text{EPP}\Leftarrow\text{OP}$: 
The intermediate execution state is one of the factors that affect the program output. Thus, the knowledge for OP covers that of PSP. 
\end{enumerate}

According to the above descriptions, we find that the knowledge required to finish the previous task is contained by that of the following task. Intuitively, the subsequent tasks are more difficult than the current task. Hence, if a model fails to correctly complete a task (e.g., fails to finish CPP) but then predicts the following tasks correctly (e.g., correct prediction of output), we consider this model behaves inconsistently in consecutive tasks.

\subsubsection{Evaluation Approach} 

We analyze the results of our Runtime Behavior Reasoning to evaluate the \newidea. Specifically, 
for the \textit{i}-th specific problem in our benchmark (i.e., full program, a specific statement in it, input, and one question to ask), we assume that the sequential results of four tasks are:
$$
\mathbf{R_{i}} = \{r_{\text{CPP}}, r_{\text{PSP}}, r_{\text{EPP}}, r_{\text{OP}}\},
$$
where $r\in\{0, 1\}$ and the number indicates whether the prediction matches the ground truth (i.e., 1) or not (i.e., 0). 
Hence, if the model is incrementally consistent and completes one task only when all its previous tasks are finished, the binary sequence $\mathbf{R_i}$ should be non-declining, i.e., 
\begin{center}\resizebox{0.98\columnwidth}{!}{
$
\mathbf{R_{i}} \in \mathbf{S}, \quad \text{where }  \mathbf{S} = \{\{1,1,1,1\},\{1,1,1,0\}, \{1,1,0,0\}, \{1,0,0,0\}\}.
$}\end{center} 

For the first example of Incremental Consistency Evaluation in Fig.~\ref{fig:framework}, the resulting sequence is $\{1, 1, 0, 0\}$ (\faCheck, \faCheck, \faTimes, \faTimes), which means that Incremental Consistency is observed in this case. However, for the second example, the result is $\{0, 1, 1, 0\}$ (\faTimes, \faCheck, \faCheck, \faTimes), so Incremental consistency is not observed.

In addition, depending on how many consecutive times consistency is maintained (i.e. $\{1,1,1,1\}$ means 4 times), we assign different weights to reward models that maintain \newidea more often. It is intuitive because it is harder to behave incrementally consistently across more sub-tasks. 

Finally, we define \textbf{\newidea Score} (IC Score) to quantitatively model the Incremental Consistency of an LLM $\mathcal{M}$. For our benchmark that contains $N$ number of $(\mathbf{P}, \mathbf{X}, I, V)$ pairs, IC Score can be computed as:
$$
\text{IC Score} = \frac{100}{N} \sum_{i=1}^{N} \text{IC Score}_i,
$$

$$
\text{IC Score}_i =
\begin{cases}
\dfrac{1}{2^{j-1}}, & \text{if } \mathbf{R}_i = \mathbf{S}_j, j \in \{1,2,3,4\} \\
0, & \text{otherwise}
\end{cases}
$$

The above formula indicates that weighted scores are given based on the number of times the model maintains \newidea. 
The higher the IC score, the higher \newidea of the model's behavior.
 Specifically, for a model's results of a problem:
\begin{enumerate}[i)]
    \item If the answers are completely correct, it gains a full score. 
    \item If the answers are partially correct and \newidea is observed, the model gains a partial score.
    \item For other cases (e.g., partially correct, but \newidea is not observed), the model gets a zero score.
\end{enumerate} 

\subsection{\coloruline[bctext]{Benchmark Construction}}
\label{subsec:benchmark}

As Fig.~\ref{fig:framework} illustrates, our framework utilizes existing executable benchmarks to evaluate the code reasoning ability of LLMs. We introduce how to adapt these base benchmarks into our framework in two steps: (i) Runtime Behavior Extraction; (ii) Problem Construction and Filtering:

\subsubsection{Runtime Behavior Extraction} 
Our evaluation framework requires code models to predict intermediate information during code execution, thus, we need to extract the runtime information as the ground truth of the problem. We use the provided test case to execute the corresponding canonical solution to ensure the correctness of program and input. During the execution, we implement the customized program tracer to record (i) the statement being executed with its number of lines and (ii) the current program state (i.e., local variables) for each execution step. Thus, when the execution of code terminates, we can acquire an ordered sequence of the runtime behavior we need for the evaluation.

\subsubsection{Problem Construction and Filtering} 
We construct our problem for each task with the extracted information. As there could be a large number of combinations of different runtime behavior and input (e.g., lots of variables in the program state of a specific time step), we design several filtering rules to select reasonable and representative ones:

\phead{CCP and EPP.} 
In these two tasks, we focus on whether and when a statement is being executed. As the actual execution sequence of statements could be very long for loops, we analyze the control flow graph and break the program into several blocks. We prioritize the last statement in a block as it leads to various new blocks and is generally more difficult to reason about.

\phead{PSP.} 
This task challenges an LLM to predict the type and value of a variable.
For \newname, we inspect the code and focus on the following types of statements: 

\begin{enumerate}
\renewcommand\labelenumi{\roman{enumi})}
\renewcommand\theenumi\labelenumi

    \item \textbf{Assignment.} We extract the variable(s) at the left-hand side for assignment statements.  The possible types of variables are identifier (i.e., ordinary variables like ``\verb|x|''), subscript (i.e., array slices like ``\verb|x[0]|''), and attribute (i.e., fields like ``\verb|x.y|''). In most cases, we extract identifiers. Note that some naive assignments such as \verb|a = 0| or \verb|l = []| are skipped, but we keep statements like \verb|a += 1| for the change of variable value.
    
    \item \textbf{Return statement.} In return statements, we extract the variables in the returned object if local variables are returned. If the returned object is a constant value, we will select the ``nearest'' variable, i.e., the last variable that is not constant.
    
    \item \textbf{Others.}  For other situations, if any variable after the current line is changed, we extract a changed variable based on the priority of ``new variable $>$ changed variables $>$ changed attributes (i.e., \verb|self.xxx|)''. Not all variables are used because we prefer variables that have closer logical relationship with other tasks (e.g., EPP). We ignore objects of non-serializable classes or complex structures (e.g., ``\verb|self|'' objects), as it is challenging to convert them to canonical string representations and compare ground truth with an LLM's output. 
\end{enumerate}

\phead{OP.} 
In output prediction task, we follow CRUXEval~\cite{gu2024cruxeval} and utilize the assertion statements in the test cases of the base benchmarks. For base benchmarks such as ClassEval~\cite{classeval}, where one test case contains multiple assertions, we use all the assertions. Specifically, we replace the right operands in the assertions with question marks (``\verb|??|''), and challenge models to predict the masked values.

After the above screening, we combine the results to obtain the final adapted benchmark, ensuring that the dataset used for each task is consistent.

\section{Experimental Setup}
\label{sec:experimental}

\subsection{Base Benchmarks}
\label{sec:base}

In our experiments, we first need to obtain the runtime behavior of code such as program state, thus base benchmarks should be executable and equipped with test cases. Moreover, we would like to experiment with diverse types of data (e.g., different programming scenarios). 
Therefore, we utilize existing code generation benchmarks as the basis for code reasoning evaluation.
Typically, code generation benchmarks can be categorized into two types: competition-level ones (i.e., with standalone functions)~\cite{chen2021evaluating, mbpp} and context-aware ones (i.e., with more code context like class and file dependencies) \cite{classeval, deveval}. 
Considering the diversity of base benchmarks, we choose HumanEval as a representative of competition-level benchmarks and ClassEval as a representative of context-aware benchmarks.

\begin{itemize}
    \item \textbf{HumanEval}~\cite{codex} is a popular competition-level benchmark for code generation. It consists of 164 hand-written Python programming problems and needs models to solve the problem given the function signature and docstring.
    \item \textbf{ClassEval}~\cite{classeval} is a class-level hand-written code generation benchmark. ClassEval provides different programming scenarios (e.g., incremental generation) and topics (e.g., management systems and database operations).
\end{itemize}

These benchmarks are widely used in previous research~\cite{liu2024codemind,zhang-etal-2023-self,hou2023large}. We show the statistics of the adapted benchmarks in Table~\ref{table:base}. In addition to our selection, we highlight that our framework is applicable for other similar code generation benchmarks.

\begin{table}
\centering
\caption{Statistics of our dataset.} 
\label{table:base}
\resizebox{0.9\columnwidth}{!}{%
\tabcolsep=10pt
\begin{tabular}{@{}l|c@{}}
\toprule
\multicolumn{1}{c|}{\textbf{Description}} & \textbf{Number} \\ \midrule
\# of Problems                       & 3152            \\
\# of Avg. Tokens in Programs                & 408.3            \\
\# of Avg. Tokens in Selected Statements     & 14.0           \\ \bottomrule
\end{tabular}
}
\end{table} 

\begin{table}
\centering
\caption{
Features of Studied LLMs. 
``FD'': Foundation (code) models. 
``IF'': Supporting instruction following. 
``OS'': Open-source models.
} 
\label{table:studied}
\resizebox{1.00\columnwidth}{!}{%
\tabcolsep=3pt
\begin{tabular}{@{}c|c|l|ccccc@{}}
\toprule
\textbf{Category} & \textbf{Series} & \multicolumn{1}{c|}{\textbf{Model Name}} & \textbf{Size} & \textbf{FD} & \textbf{IF} & \textbf{OS} & \textbf{Time} \\ \midrule
\multirow{10}{*}{\begin{tabular}[c]{@{}c@{}}Code\\ LLMs\end{tabular}} & \multirow{5}{*}{CodeLlama} & CodeLlama-7B-Base & 7B & \ding{51} & \ding{55} & \ding{51} & 08/2023 \\
 &  & CodeLlama-7B-Python & 7B & \ding{51} & \ding{55} & \ding{51} & 08/2023 \\
 &  & CodeLlama-7B-Instruct & 7B & \ding{51} & \ding{51} & \ding{51} & 08/2023 \\
 &  & CodeLlama-13B-Instruct & 13B & \ding{51} & \ding{51} & \ding{51} & 08/2023 \\
 &  & CodeLlama-34B-Instruct & 34B & \ding{51} & \ding{51} & \ding{51} & 08/2023 \\ \cmidrule(l){2-8} 
 & \multirow{2}{*}{Magicoder} & Magicoder-CL-7B & 7B & \ding{55} & \ding{51} & \ding{51} & 12/2023 \\
 &  & Magicoder-S-CL-7B & 7B & \ding{55} & \ding{51} & \ding{51} & 12/2023 \\ \cmidrule(l){2-8} 
 & \multirow{3}{*}{StarCoder2} & StarCoder2-3B & 3B & \ding{51} & \ding{51} & \ding{51} & 02/2024 \\
 &  & StarCoder2-7B & 7B & \ding{51} & \ding{55} & \ding{51} & 02/2024 \\
 &  & StarCoder2-15B & 15B & \ding{51} & \ding{55} & \ding{51} & 02/2024 \\ \midrule
\multirow{5}{*}{\begin{tabular}[c]{@{}c@{}}General\\ LLMs\end{tabular}} & \multirow{2}{*}{GPT} & GPT-3.5-Turbo & - & - & \ding{51} & \ding{55} & 01/2024 \\
 &  & GPT-4-Turbo & - & - & \ding{51} & \ding{55} & 01/2024 \\ \cmidrule(l){2-8} 
 & Mistral & Mistral-7B-Instruct & 7B & \ding{55} & \ding{51} & \ding{51} & 01/2024 \\ \cmidrule(l){2-8} 
 & \multirow{2}{*}{Gemma} & Gemma-7B-It & 7B & \ding{51} & \ding{51} & \ding{51} & 02/2024 \\
 &  & Gemma-2B-It & 2B & \ding{51} & \ding{51} & \ding{51} & 02/2024 \\ \bottomrule
\end{tabular}
}
\end{table}

\subsection{Studied Code LLMs}
\label{sec:studied}

To study the reasoning capabilities of diverse code LLMs, we curate a selection of models with a variety of distinctions. 
Specifically, we mainly consider these dimensions of them:
(1) general or code specified (e.g., GPT-4-Turbo~\cite{gpt4} v.s. CodeLlama~\cite{codellama});
(2) scale of parameters;
(3) open-source or closed-source (e.g., GPT-3.5-Turbo~\cite{gpt4} v.s. Mistral-7B-Instruct~\cite{mistral})
(4) foundation or further fine-tuned (e.g., CodeLlama v.s. Magicoder-CL~\cite{magicoder});
(5) instruct following or not (e.g., CodeLlama-7B-Base v.s. its instruct version);
(6) open-source or not; and (7) release time.

As a result, considering these dimensions we select several state-of-the-art code LLMs that have been applied to various code related tasks~\cite{yang2024large,shypula2024learning,wei2023copiloting,chen2024nlperturbator}.
Table~\ref{table:studied} presents detailed features of them and we can see that full consideration of the diversity of models across various features is taken to enhance the generalization of our study.

\subsection{Prompt Design}
\label{subsec:prompt} 

\begin{figure}[t]
\centerline{\includegraphics[width=1.0\linewidth]{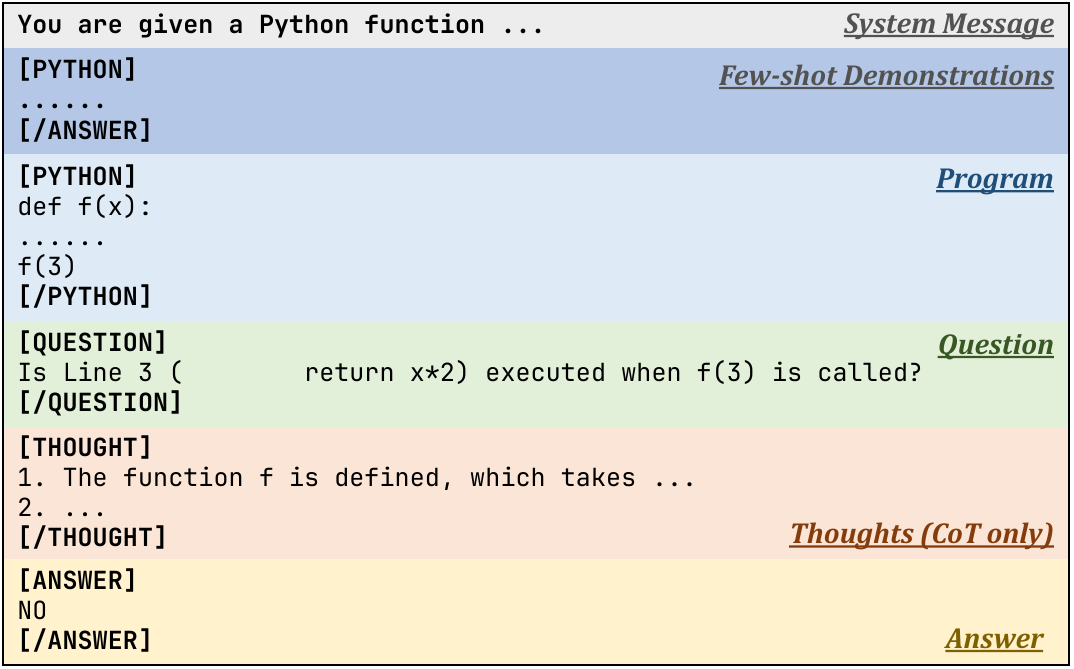}}
\caption{The prompt template for our empirical study. Note that the ``Thoughts'' part is used only we leverage Chain-of-Thought (CoT)~\cite{wei2022chain} prompting.}
\label{fig:prompt}
\end{figure}

In our work, we utilize prompting to evaluate code LLMs with our code reasoning tasks. 
We refer to a recent study~\cite{gu2024cruxeval} on code reasoning and design our prompt templates as illustrated in Fig.~\ref{fig:prompt}. 
For classic few-shot prompting, our prompt template consists mainly of five parts, including the system message, few-shot demonstrations, program, question, and answer. If Chain-of-Thought (CoT)~\cite{wei2022chain} prompting is utilized, the thoughts are added into the prompt template as well as the examples in it. 
Both few-shot prompting and CoT prompting are widely applied in various tasks~\cite{icse24_code_suggestion_rag,song2023comprehensive,wei2022chain}, including reasoning tasks~\cite{gu2024cruxeval}.

\subsection{Implementation Details} 
\label{sec:implementation}

\phead{Access of Models and Base Benchmarks.}
For open-source models such as CodeLlama, we use the corresponding official releases available on HuggingFace~\cite{hf}. For closed-source models (e.g., GPT-3.5-Turbo and GPT-4-Turbo), we invoke the OpenAI API~\cite{api} to access them. 
Two benchmarks (i.e., HumanEval and ClassEval) are also publicly available on HuggingFace~\cite{hfhumaneval,hfclasseval}.
To help replicate our research, we list detailed information, such as model IDs and URLs in our replication package~\cite{replication}.

\phead{Environment.} We run experiments on a Linux server with 8 NVIDIA A800 GPUs. For open-source LLMs, we deploy a local API server based on vLLM~\cite{kwon2023efficient} which is a unified library for LLM serving and inference. All models are not quantized and we use their original precisions.

\phead{Configurations.} 
Temperature can control the randomness in the generated results of models~\cite{codex}.
Specifically, we follow Gu et al.~\cite{gu2024cruxeval} and set the temperature to 0.8. For tasks with direct prompting, we set the maximum length of generated tokens to 256, while for tasks with CoT prompting, we set it to 1024. For the rest of the parameters, we use the default settings in vLLM, to ensure a fair comparison. To obtain reliable results, experiments for all open-source models with few-shot prompting are repeated five times, and we report the mean and standard deviation values in Section~\ref{sec:results}. We do not repeat experiments for closed-source models like the GPT series due to a limited budget. 


\begin{table*}[tb]
\vspace{-0.3cm}\caption{
Results for Runtime Behavior Reasoning and \newidea Evaluation (RQ1 \& 2). 
``CCP'', ``PSP'', ``EPP'', and ``OP'': four tasks of Runtime Behavior Reasoning; 
``Avg'': the average accuracy score of four tasks. \\
We report the results in the form of ``\textbf{mean±standard deviation}" except for two GPT models because of budget limit.
}
\centering
\label{table:rq1}
\resizebox{\linewidth}{!}{
\tabcolsep=8pt
\begin{tabular}{@{}l|cccccc|c@{}}
\toprule
\multicolumn{1}{c|}{\multirow{2}{*}{\textbf{Model}}} & \multicolumn{2}{c}{\textbf{CCP}} & \textbf{PSP} & \textbf{EPP} & \textbf{OP} & \textbf{Acc. Avg.} & \textbf{IC} \\ \cmidrule(l){2-8} 
\multicolumn{1}{c|}{} & Acc. (\%) & F1 & Acc. (\%) & Acc. (\%) & Acc. (\%) & (\%) & Score \\ \midrule
CodeLlama-7B-Base & 54.3±0.5 & 56.1±0.5 & 25.0±0.5 & 5.6±0.3 & 58.2±1.6 & 35.8 & 4.0±0.2 \\
CodeLlama-7B-Python & 55.5±0.7 & 62.7±0.6 & 31.3±1.0 & 8.7±0.7 & 62.3±1.0 & 39.4 & 4.8±0.2 \\
CodeLlama-7B-Instruct & 55.6±0.9 & 47.2±1.2 & 25.1±0.5 & 10.8±0.2 & 62.6±0.8 & 38.5 & 4.1±0.1 \\
CodeLlama-13B-Instruct & 61.0±0.7 & 66.4±0.6 & 32.5±0.4 & 14.4±0.4 & 64.5±1.1 & 43.1 & 6.6±0.3 \\
CodeLlama-34B-Instruct & 61.5±0.5 & 70.1±0.4 & 47.5±0.6 & 29.2±0.4 & 65.9±1.1 & 51.0 & 11.8±0.3 \\ \midrule
StarCoder2-3B & 54.8±0.7 & 58.2±0.5 & 29.0±0.7 & 6.5±0.5 & 58.8±0.8 & 37.3 & 4.3±0.3 \\
StarCoder2-7B & 55.1±0.7 & 63.8±0.6 & 34.2±0.7 & 5.0±0.4 & 63.9±0.8 & 39.6 & 4.2±0.3 \\
StarCoder2-15B & 58.9±0.8 & 64.6±0.8 & 43.5±0.3 & 28.0±0.5 & 71.5±1.1 & 50.5 & 10.7±0.4 \\ \midrule
Magicoder-CL & 58.7±1.4 & 61.2±1.8 & 30.1±0.5 & 15.5±1.1 & 60.4±1.4 & 41.2 & 6.2±0.3 \\
Magicoder-S-CL & 60.3±1.1 & 69.9±0.8 & 31.4±0.4 & 9.8±0.4 & 62.3±1.2 & 40.9 & 6.0±0.2 \\ \midrule
Gemma-2B-It & 52.7±0.4 & 31.0±0.6 & 13.5±0.5 & 7.3±0.5 & 43.9±1.5 & 29.3 & 5.5±0.2 \\
Gemma-7B-It & 66.3±0.3 & 75.2±0.1 & 32.1±0.1 & 8.4±0.4 & 57.9±0.7 & 41.2 & 6.9±0.2 \\ \midrule
Mistral-7B-Instruct & 69.5±0.2 & 75.9±0.2 & 35.2±0.3 & 35.8±0.4 & 51.5±0.7 & 48.0 & 16.3±0.3 \\ \midrule
GPT-3.5-Turbo & 61.8 & 64.0 & 51.6 & 48.6 & 60.7 & 55.7 & 20.6 \\
GPT-4-Turbo & \textbf{88.4} & \textbf{89.8} & \textbf{71.4} & \textbf{57.7} & \textbf{82.6} & \textbf{75.0} & \textbf{42.5} \\ \midrule
\multicolumn{1}{c|}{\textbf{Average}} & 61.0 & 63.7 & 35.6 & 19.4 & 61.8 & 44.4 & 10.3 \\ \midrule
CodeLlama-7B-Instruct (CoT) & 57.5 & 59.2 & 33.4 & 21.4 & 55.8 & 42.2 & 7.5 \\ \bottomrule
\end{tabular}
}\vspace{-0.3cm}
\end{table*}

\section{Results}
\label{sec:results}

In this section, we discuss the results of our empirical study on \newname by answering two research questions:
\begin{itemize}
    \item \textbf{RQ1}: How do LLMs perform on Runtime Behavior Reasoning?
    

    \item \textbf{RQ2}: How do LLMs perform on \newidea evaluation?
\end{itemize}

\subsection{RQ1: Performance of Runtime Behavior Reasoning}
\label{subsec:rq1}

Table~\ref{table:rq1} shows the detailed results of Runtime Behavior Reasoning. 
All models are evaluated with few-shot prompting except for special annotated ones (i.e., CodeLlama-7b-Instruct (CoT)). Below, we discuss the results from different aspects.

\phead{Overall Performance.} 
Overall, we find that the performance of different LLMs presents a large variation, and GPT-4-Turbo shows superior performance in reasoning about program execution. For example, GPT-4-Turbo achieves the best results in all metrics of Runtime Behavior Reasoning, and its average accuracy outperforms the second best (i.e., 55.7\% of GPT-3.5) by a large margin (i.e., an absolute improvement of 19.3\%). However, the overall performance of open-source models is not high and the best performer among them (i.e., CodeLlama-34B-Instruct) only achieves a level close to that of GPT-3.5 (i.e., 51.0\% v.s. 55.7\%) in terms of average accuracy.

\phead{Task.} 
Runtime Behavior Reasoning consists of four distinct evaluation tasks, i.e., CCP, PSP, EPP, and OP, and the performance varies among different tasks. 
For example, all models achieve an accuracy of more than 50\% in OP (i.e., Output Prediction), while only about half of them (i.e., 8 out of 15) can provide the correct answers for more than 10\% problems in EPP (i.e., Execution Path Prediction). Hence, according to the average score of tasks, the performance distribution may suggest that EPP is the most challenging task and OP is relatively easy among them.

\phead{Size and Category.} 
In general, we observe that for models within the same family, the variant with larger size of parameters shows better performance in Runtime Behavior Reasoning. 
In the case of the CodeLlama-instruct series, as the number of parameters increases (i.e., 7B $\rightarrow$ 34B), the accuracy of EPP has a relative improve by over 100\% (i.e., 14.4\% $\rightarrow$ 29.2\%). 
Meanwhile, smaller models like StarCoder2-3B can also outperform larger models such as CodeLlama-7B-Instruct in terms of average accuracy (i.e., 37.3\% v.s. 35.8\%). 
The StarCoder2 series utilizes varied architectures and training datasets compared to CodeLlama. This may demonstrate that apart from parameter size, the model architecture and training strategy also play an important role in code reasoning ability.
We also find that code LLMs do not exhibit an obviously leading advantage over general LLMs of the same size.

\phead{Training Strategy.} 
As shown in Table~\ref{table:rq1}, we conduct experiments on three variants of CodeLlama (i.e., base, instruct, and Python) of the same 7B size. Compared to the base model, the ``instruct'' variant that leverages instruction tuning techniques brings gains in the code reasoning ability (i.e., Avg. Acc. from 35.8\% to 38.5\%), which may reflect the relationship between understanding instructions and reasoning program. Meanwhile, since our base benchmarks are all in Python, additional training with Python corpora (i.e., CodeLlama-Python) leads to an improvement of performance (i.e., an absolute improvement of 0.9\%). 
In addition, we note that although further fine-tuning applied to the Magicoder series improves their performance in code generation~\cite{magicoder}, the improvement in code reasoning ability is relatively limited compared to their foundation model CodeLlama-7B-Python. 
This may indicate that the training strategies they utilize are not well suited for the reasoning tasks in our evaluation.

\phead{Prompting Strategy.} 
The last row of Table~\ref{table:rq1} presents the performance of CodeLlama-7B-Instruct with CoT prompting. 
Compared to the model with few-shot prompting, the performance of CPP, PSP, and EPP receives varying degrees of improvement. For instance, the EPP accuracy with CoT is improved from 10.8\% to 21.4\%, surpassing the performance of the larger 13B model (i.e., 14.4\%). It may demonstrate the effectiveness of presenting how to reason about a piece of code step by step. 
However, CoT prompting fails to improve its OP performance, with an absolute decrease of 6.8\%. This may result from the wrong thought chain that the model generates for the whole program and eventually leads to the mistake.

\greyboxb{Summary for RQ1:} {
Models with different features (e.g., size and training strategy) exhibit notable disparities in performance on Runtime Behavior Reasoning. 
Overall, GPT-4-Turbo demonstrates a clear advantage over other models in all four tasks in our setting.
}

\subsection{RQ2: \newidea Evaluation}
\label{subsec:rq2}

\begin{figure}[t]
\centerline{\includegraphics[width=1.0\columnwidth]{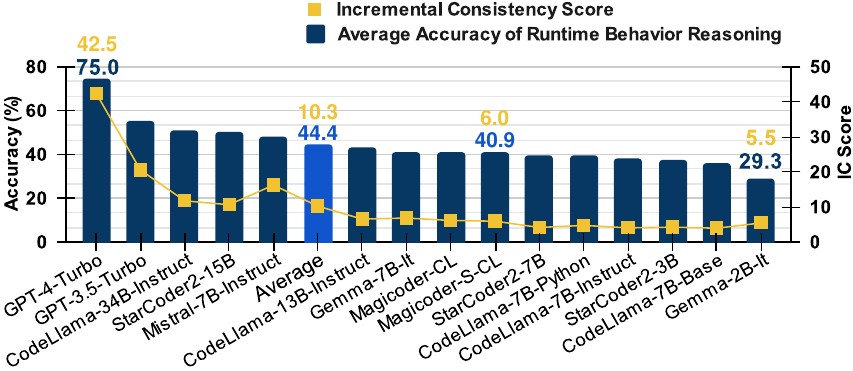}}
\caption{
Average Accuracy of Runtime Behavior Reasoning and \newidea Score for different models, sorted in descending order according to average accuracy.
}
\label{fig:isc}
\end{figure}

Fig.~\ref{fig:isc} shows the sorted average accuracy for Runtime Behavior Reason for different models, with an additional line indicating the IC scores. A detailed information of mean IC scores and their standard deviations are also reported in Table~\ref{table:rq1}.

\phead{Overall Performance.} 
We find that the majority of LLMs exhibit a low level of \newidea with scores below 20, which highlights the inconsistency in model behavior across the four tasks of Runtime Behavior Reasoning.
Among all of the models, GPT-4-Turbo stands out with the highest IC score of 42.5, even more than double that of the second place GPT-3.5 (i.e., 20.6). 
Given that GPT-4-Turbo achieves the best results in both Runtime Behavior Reasoning and \newidea Evaluation (i.e., 75.0\% of Avg. Acc. and 42.5 of IC score), we believe that it has both superior ability in program reasoning and a high level of \newidea across sequentially related tasks in our evaluation.

\phead{Trend Between IC and Runtime Behavior Reasoning.} 
As illustrated in Fig.~\ref{fig:isc}, we find that there is an approximately similar trend between the model's average accuracy and its IC score.
For example, compared to CodeLlama-7B-Instruct, its larger version CodeLlama-34B-Instruct has a noticeable higher average accuracy (i.e., 38.5\% v.s. 51.0\%) and IC score (i.e., 4.1 v.s. 11.8). 
However, this pattern does not hold for all models. The performance of the general LLM Mistral-7B is not as good as that of CodeLlama-34B in terms of average accuracy (i.e., 48.0\% v.s. 51.0\% in Runtime Beheavior Reasoning), but performs better on \newidea (i.e., 16.3 v.s. 11.8).


\phead{Others.} 
Similar to the results of RQ1, we observe that code LLMs do not significantly outperform general LLMs in the IC evaluation. 
For example, the IC score of Gemma-2B-It (i.e., 5.5) is higher than code LLMs trained with more code corpora like CodeLlama-7b-Instruct and StarCoder2-7B (i.e., 4.1 and 4.2). This phenomenon may suggest that more code data cannot help LLMs reason programs better and maintain their \newidea.
In addition, CoT prompting for CodeLlama-7B-Instruct leads to a great increase in its IC Score (i.e., 4.1 $\rightarrow$ 7.5), and this improvement of IC Score may benefit from explicit problem solving steps.

\greyboxb{Summary for RQ2:}{ 
In code reasoning tasks, most LLMs behave inconsistently and their average accuracy is not entirely associated with IC. 
GPT-4-Turbo achieves an IC Score of as high as 42.5, surpassing other models by a large margin (i.e., more than 21.9 absolute improvements).
}


\section{Discussion}
\label{sec:discussion}

\subsection{Case Study}
\label{subsec:case}

\begin{figure}[t]
\centerline{\includegraphics[width=0.8\linewidth]{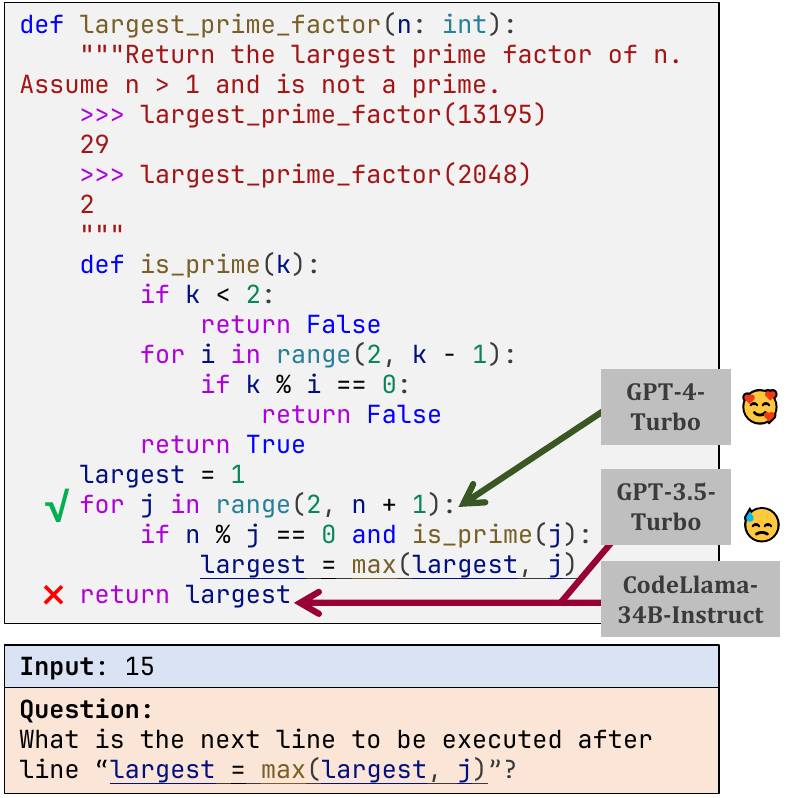}}
\caption{
A tricky problem of EPP from HumanEval/59~\cite{codex}. 
The prediction of GPT-4-Turbo is correct and the other two models (i.e., GPT-3.5-Turbo and CodeLlama-34B-Instruct) fail to finish it. The problem description is simplified for a concise presentation.}
\label{fig:case1}
\end{figure}

Fig.~\ref{fig:case1} shows a case from the problem of EPP. Given a Python function that aims to return the largest prime factor of parameter \verb|n| and input $15$, this problem requires the model to predict the next statement to be executed after an assignment statement (i.e., \verb|largest = max ...|). 
Here we select three models that show competitive performance in EPP including GPT-4-Turbo, GPT-3.5-Turbo, and CodeLlama-34B-Instruct. For GPT-3.5-Turbo and CodeLlama-34B-Instruct, the prediction is the last ``return'' statement of this function; while GPT-4-Turbo chooses the above ``for'' loop as its prediction. We mark a happy emoji to show that GPT-4-Turbo makes the right choice, and the other two models fail to predict it correctly. The explanation is that if the assignment statement is executed, the possible value of $j$ can only be 3 or 5 ($n + 1 = 16$), which means that the loop will continue and the ``return'' statement is not the next executed line. 

In our framework and the adapted benchmark, there are many problems like this that have no obvious answer and are challenging. If the model is not capable of reasoning about its inherent logic of execution, it can easily be misled and give the most ``look-alike'' answer (i.e., the ``return'' statement in this case), indicating that our framework can effectively measure the code reasoning capability of LLMs and present the discrimination of them. We discuss more cases in the appendix, which can be accessed in our replication package\footnote{\url{https://r-eval.github.io}}.

\subsection{Unsatisfactory Performance of Code Reasoning} 



According to evaluation results, we observe that many models perform poorly in Runtime Behavior Reasoning and \newidea Evaluation. In particular, even the best performer GPT-4-Turbo only achieves an IC score of 42.5, reflecting the limitation of current models in maintaining consistency in sequential-related tasks, and there is still a long way to go to make the LLMs perform code reasoning.

One potential reason is that the current LLM might not understand the program execution behavior. While it is convenient to obtain source code from open-source platforms (e.g., GitHub), there is relatively less data available regarding code execution behavior, because running the program and collecting its runtime information require the corresponding development environments and test suite. 
Therefore, if the model is not familiar with the knowledge related to runtime behavior, it may not perform well in code reasoning tasks.


\begin{table}[t]
\centering
\caption{Pearson Correlation Coefficient Matrix of the Results of Runtime
Behavior Reasoning (RBR), Incremental Consistency (IC), and HumanEval (HE)}
\label{tab:corr}
\renewcommand\arraystretch{1.4}
\begin{tabular}{|c|ccc|}
\hline
\textbf{Pearson Correlation} & \textbf{RBR}                  & \textbf{IC}                   & \textbf{HE}                   \\ \hline
\textbf{RBR} & \cellcolor[HTML]{D16D6A}1.000 & \cellcolor[HTML]{D78280}0.940 & \cellcolor[HTML]{EAC0BF}0.772 \\
\textbf{IC}                  & \cellcolor[HTML]{D78280}0.940 & \cellcolor[HTML]{D16D6A}1.000 & \cellcolor[HTML]{F0D1D1}0.724 \\
\textbf{HE}                  & \cellcolor[HTML]{EAC0BF}0.772 & \cellcolor[HTML]{F0D1D1}0.724 & \cellcolor[HTML]{D16D6A}1.000 \\ \hline
\end{tabular}
\end{table}

\subsection{Correlation between Code Reasoning \& Code Generation}

We utilize the experimental results and study the correlation between code reasoning and code generation, i.e., whether an LLM that performs well in code generation could exhibit equally strong abilities in code reasoning. 

Table~\ref{tab:corr} presents the Pearson correlation coefficient matrix of Runtime Behavior Reasoning (Avg. Acc.), Incremental Consistency Score and HumanEval (\textit{pass@}$1$ rate). According to the matrix, we find that there is a strong positive correlation (i.e, the Pearson correlation coefficients are larger than 0.7) among code generation (i.e., HumanEval) and code
reasoning (i.e., Runtime Behavior Reasoning and Incremental Consistency). However, the correlation between code reasoning and
code generation is relatively lower than that between two reasoning tasks internally (i.e., 0.724 and 0.772 v.s. 0.940), which indicates that models with similar code generation abilities may vary a lot in code reasoning.
As the correlation may help researchers increase the understanding of code LLMs and improve the models' code generation
and reasoning abilities, future research could investigate such correlation in depth.

\subsection{Threats to Validity}  

\phead{Internal Threats.}
To construct the adapted benchmark for different sub-tasks of Runtime Behavior Reasoning, we manually establish some rules to select appropriate statements and variables for the evaluation. However, our selection criteria may not effectively represent the runtime state of the program. 
To mitigate this threat, we take some measures to determine problem settings that are representative and challenging, based on the characteristics of different tasks. For instance, we choose the last statement in the control flow (i.e., for EPP) and variables that are modified after the execution (i.e., for PSP). These measures help us to reasonably assess the model's capability to reason about code and provide meaningful differentiation. 
For Runtime Behavior Reasoning, we select four dimensions of the intermediate state of program execution which are widely applied in previous research~\cite{ding2024traced, tufano2023predicting}. These four tasks are proven to effectively evaluate the code reasoning capability of code LLMs~\cite{tufano2023predicting,ding2024traced,liu2024codemind}, and are appropriate for \newidea Evaluation for their unique sequential relationship. 
However, there are still some dynamic features such as memory allocation and exception handling which may help measure code models, and we have not explored yet. 
Further research could consider exploring the potential for LLMs to reason about other dynamic program features and extend \newname to more scenarios.

\phead{External Threats.}
In the empirical study for our evaluation framework, the results are restricted to the specific collection of code models and base benchmarks. 
To mitigate this threat, we choose representative code LLMs considering several standards including their scale, popularity, and training strategy; For the base benchmarks applied, two benchmarks (i.e., HumanEval~\cite{codex} and ClassEval~\cite{classeval}) are distinct from evaluation fashion and programming scenarios, as described in Section~\ref{subsec:benchmark}. With the above efforts, the experimental results are expected to be sustained in more circumstances. 

\section{Conclusion and Future Work}
\label{sec:conclusion}

In this paper, we propose \newname, a comprehensive framework for evaluating the code reasoning capability of code LLMs. 
Our framework consists of two evaluation components including Runtime Behavior Reasoning and \newidea Evaluation: 
We conduct a large-scale empirical study on several popular LLMs and two widely used base benchmarks.  
Our empirical results show that the majority of LLMs we evaluate show unsatisfactory performance in both Runtime Behavior Reasoning and \newidea Evaluation. To improve the code reasoning capabilities of LLMs,  future works can explore: 

\phead{Training with Execution Behavior.} 
One reason why large models may struggle with code reasoning is possibly due to a lack of knowledge related to program execution. 
Although some general fine-tuning approaches are applied to code LLMs~\cite{magicoder}, they fail to improve the code reasoning capabilities.  
Given the demonstrated effectiveness of training models with execution behavior in improving performance in a range of downstream tasks~\cite{ding2024traced, liu2023code}, it is reasonable to expect that LLMs would also derive benefits from such a process.

\phead{Improving Prompting Strategy.} 
Our evaluation results demonstrate the effectiveness of CoT prompting in code reasoning tasks. 
Apart from CoT, other prompting techniques that have been proven effective in NL reasoning tasks (such as Tree-of-Thoughts~\cite{yao2024tree}) may also be applicable to code reasoning tasks. 
Besides, the prompting approach tailored for reasoning about program execution also warrants investigation.

\section*{Acknowledgments} 
This research is supported by the Ningbo Natural Science Foundation (No. 2023J292). It is also supported by the advanced computing resources provided by the Supercomputing Center of Hangzhou City University.

\bibliographystyle{IEEEtran}
\bibliography{references}

\end{document}